\documentclass[twocolumn]{aastex63}
\usepackage{gensymb}
\usepackage{hyperref}

\accepted{November 18, 2022}
\submitjournal{PSJ}

\shorttitle{Io's surface from UV-visible spectroscopy}
\shortauthors{Trumbo et al.}

\graphicspath{{./}{figures/}}

\begin{document}

\title{Spectroscopic mapping of Io's surface with HST/STIS: SO$_2$ frost, sulfur allotropes, and large-scale compositional patterns}

\correspondingauthor{Samantha K. Trumbo}
\email{skt39@cornell.edu}

\author[0000-0002-0767-8901]{Samantha K. Trumbo}
\affiliation{Cornell Center for Astrophysics and Planetary Science, Cornell University, Ithaca, NY 14853, USA}

\author[0000-0002-7451-4704]{M. Ryleigh Davis}
\affiliation{Division of Geological and Planetary Sciences, California Institute of Technology, Pasadena, CA 91125, USA}

\author[0000-0002-9544-0118]{Benjamin Cassese}
\altaffiliation{Now at Department of Astronomy, Columbia University, New York, NY 10027 USA}
\affiliation{Division of Geological and Planetary Sciences, California Institute of Technology, Pasadena, CA 91125, USA}

\author[0000-0002-8255-0545]{Michael E. Brown}
\affiliation{Division of Geological and Planetary Sciences, California Institute of Technology, Pasadena, CA 91125, USA}

\begin{abstract}
Io's intense volcanic activity results in one of the most colorful surfaces in the solar system. Ultraviolet and visible-wavelength observations of Io are critical to uncovering the chemistry behind its volcanic hues. Here, we present global, spatially resolved UV-visible spectra of Io from the Space Telescope Imaging Spectrograph on the Hubble Space Telescope (HST), which bridge the gap between previous highly resolved imagery and disk-integrated spectroscopy, to provide an unprecedented combination of spatial and spectral detail. We use this comprehensive dataset to investigate spectral endmembers, map observed spectral features associated with SO$_2$ frost and other sulfur species, and explore possible compositions in the context of Io surface processes. In agreement with past observations, our results are consistent with extensive equatorial SO$_2$ frost deposits that are stable over multi-decade timescales, widespread sulfur-rich plains surrounding the SO$_2$ deposits, and the enrichment of Pele's pyroclastic ring and the high-latitude regions in metastable short-chain sulfur allotropes.
\end{abstract}


\section{Introduction} \label{sec:intro}
Jupiter's moon Io is the most volcanically active body in the solar system. Explosive and effusive eruptions driven by the intense tidal heating associated with its forced eccentric orbit \citep{Peale1979} cause constant volcanic resurfacing \citep[e.g][]{Williams2007,Geissler2003,Geissler2007}, making Io the body with the youngest surface in the solar system as well. Its dynamic surface and coupled atmosphere are dominated by volcanic products, which also feed neutral clouds and a plasma torus beyond Io within Jupiter's extensive magnetosphere \citep[e.g][]{Schneider2007}, resulting in the delivery of Iogenic material (primarily in the form of sulfur and oxygen ions) all the way to the outer, icy Galilean satellites \citep{Johnson2004}. On Io, these volcanic products give rise to an extremely colorful surface, featuring dark lava lakes and flows, vividly red rings of volcanic plume fallout, widespread plains of yellow, orange, and brown, and regions coated in bright sulfur dioxide (SO$_2$) frost \citep[e.g.][]{McEwen1998, Geissler1999, carlson2007io}. Ultraviolet- (UV) and visible-wavelength observations of Io's surface have been key to understanding the composition behind these colors and what they reveal about the chemistry of Io's plumes and magma reservoirs.

Prior to the \textit{Voyager} mission's first flyby of the Jupiter system, Io's unusual color and low near-UV reflectance in Earth-based imagery \citep{Harris1961, Minton1973, Murray1975} and narrow-band spectrophotometry \citep{Johnson1970, Johnson1971, Wamsteker1973, Nelson1978} were variably suggested to reflect polysulfides \citep{JohnsonMcCord1971}, elemental sulfur \citep{Wamsteker1974}, sulfur-rich salt evaporite deposits featuring irradiation-produced color centers \citep{Fanale1974, NashFanale1977, Fanale1977}, and mixtures of metastable sulfur allotropes and possibly evaporite salts \citep{Nelson1978}. \textit{Voyager's} discovery of active volcanism \citep{Morabito1979} and volcanic SO$_2$ gas \citep{Pearl1979} at Io led to the focus on plausible volcanic species and the quick identification of condensed surficial SO$_2$ \citep{Hapke1979,Fanale1979,Smythe1979} via previously unidentified infrared bands \citep{Cruikshank1978, Pollack1978} and a UV absorption maximum near 280 nm measured by the International Ultraviolet Explorer (IUE) \citep{Bertaux1979, Nelson1980}. The variegated surface colors that \textit{Voyager} revealed were suggested to reflect various volcanic sulfur products, including elemental sulfur, metastable allotropes, and possibly polysuflur oxides (PSO),  disulfur monoxide (S$_2$O), and certain sulfides, though SO$_2$ was still the only material identified with any confidence \citep{McEwen1988color, McEwen1988so2}.

Following the Voyager flybys, the existing disk-integrated spectrophotometry measurements \citep[e.g.][]{Nelson1978,McFadden1980}, which provided more spectral detail than the broadband images, but without spatial resolution, were modeled using different subsets of these possible volcanic species, with some authors favoring oxides \citep{Hapke1989}, some invoking a mixture of oxides and sodium sulfide (Na$_2$S) \citep{Nash1993}, and others favoring quenched sulfur allotropes \citep{MosesNash1991}. Io's UV-visible spectrum at the full-disk scale is characterized by two strong, but not entirely diagnostic, features---a strong absorption edge from $\sim$380 to 500 nm and a second absorption edge between 550 and 700 nm, with a plateau between them \citep[e.g.][]{JohnsonMcCord1971}. Later high-quality disk-integrated CCD spectroscopy confirmed these absorptions, but still did not differentiate between possible compositions \citep{Spencer1995}. 

Ahead of the arrival of the \textit{Galileo} spacecraft at Jupiter, \citet{Spencer1997} used the Wide-Field/Planetary Camera 2 on the Hubble Space Telescope (HST) to obtain images in filters targeting the absorption edges. \citet{Spencer1997} created spectral ratio maps that suggested that the 550-700 nm band edge was specifically associated with both the darkened high latitudes and the ring of pyroclastic deposits surrounding the extremely active Pele volcano. The authors suggested metastable tetrasulfur (S$_4$), a likely red colorant in laboratory studies of sulfur melts and S$_2$O \citep[e.g.][]{Meyer1972,Hapke1989}, as a possible explanation. Images captured with the \textit{Galileo} Solid State Imaging (SSI) system's expanded coverage into the red and near-infrared wavelengths soon confirmed a distinct red color for Pele's ring, as well as a redness to the high latitudes \citep{McEwen1998,Geissler1999}, which was originally noted from Earth by \citet{Minton1973}. Informed by the Earth-based observations, interpretation of the corresponding broadband photometry through comparison with laboratory data led to a general post-\textit{Galileo} compositional picture of yellow plains containing sulfur, bright deposits rich in SO$_2$, red materials possibly enriched in short-chain sulfur allotropes or other colorants, and dark regions of exposed volcanic silicates accounting for just 1\% of the surface \citep{Geissler1999,carlson2007io}.

Though this understanding represents the culmination of decades of observation, it still relies on inference to bridge the gap between spatially resolved images, which lack spectral detail, and full-disk UV-visible spectroscopy, which lack the necessary spatial information to link composition with surface processes. To address this problem and complete our spectral understanding of Io's colorful surface, we used HST's Space Telescope Imaging Spectrograph (STIS) to obtain the first global, spatially resolved UV-visible reflectance spectra of Io. We use this new dataset to investigate spectral endmembers, identify and map observed spectral features, and discuss compositional implications in the context of Io surface processes.

\section{Observations and Data Reduction} \label{sec:observations}
We present spatially resolved spectra of Io from 200 to 1000 nm, which we obtained with HST/STIS across 8 visits in 2020. Using the G230L, G430L, and G750L first-order spectroscopy modes, we executed simple slit-scan patterns to build up complete spatial coverage of Io's disk during each visit. The G230L UV observations used the STIS MAMA detector, while the longer-wavelength gratings were used in combination with the STIS CCD. For G230L, we stepped the 0.2$^{\prime\prime}$ slit across Io in 0.1$^{\prime\prime}$ increments, which provided high-quality spectra from 200 to 300 nm at an effective spatial resolution of 0.1$^{\prime\prime}$ in the across-slit direction (corresponding to $\sim$330 km at the sub-observer point). For G430L and G750L, we used the 0.1$^{\prime\prime}$ slit and stepped in 0.06$^{\prime\prime}$ increments, providing approximately twice the across-slit spatial resolution in the 300--1000 nm range. Details concerning the dates, geometries, and exposure times associated with each visit and slit scan are given in Table \ref{table:obs}.

\begin{table*}[ht]
\begin{center}
\caption{Table of HST/STIS Observations\label{table:obs}}
\begin{tabular}{cccccccccc}
\hline\\[-4mm] \hline
Date&Time&Central&Central&Angular&Gratings&Slit&Pixel&Resolving&Exposure\\
(UT)&(Start/End)&Lon.&Lat.&Diameter&Used&Width\footnote{The effective across-slit angular resolution is finer by roughly a factor of two for all settings because we overlapped subsequent slit positions in our stepping patterns, as described in Section 2.}&Scale&Power&Time (s)\\ \hline
2020 Mar 25 & 02:59/03:31 &  225$\degree$W & -1.56$\degree$ & 0.93$^{\prime\prime}$ & G230L & 0.2$^{\prime\prime}$ & 0.025$^{\prime\prime}$ & 500 & 167\\
2020 Aug 7 & 10:31/11:03 &  45$\degree$W & -1.33$\degree$ & 1.20$^{\prime\prime}$ & G230L & 0.2$^{\prime\prime}$ & 0.025$^{\prime\prime}$ & 500 & 167\\
2020 Aug 11 & 09:50/10:22 &  134$\degree$W & -1.34$\degree$ & 1.19$^{\prime\prime}$ & G230L & 0.2$^{\prime\prime}$ & 0.025$^{\prime\prime}$ & 500 & 167\\
2020 Aug 21 & 03:20/03:52 &  314$\degree$W & -1.34$\degree$ & 1.16$^{\prime\prime}$ & G230L & 0.2$^{\prime\prime}$ & 0.025$^{\prime\prime}$ & 500 & 167\\
2020 Aug 18 & 11:48/12:22 &  136$\degree$W & -1.34$\degree$ & 1.17$^{\prime\prime}$ & G430L/G750L & 0.1$^{\prime\prime}$ & 0.05$^{\prime\prime}$ & 500 & 7/6\\
2020 Aug 28 & 05:23/05:57 &  317$\degree$W & -1.35$\degree$ & 1.14$^{\prime\prime}$ & G430L/G750L & 0.1$^{\prime\prime}$ & 0.05$^{\prime\prime}$ & 500 & 7/6\\
2020 Sep 10 & 22:21/22:55 &  226$\degree$W & -1.34$\degree$ & 1.10$^{\prime\prime}$ & G430L/G750L & 0.1$^{\prime\prime}$ & 0.05$^{\prime\prime}$ & 500 & 7/6\\
2020 Sep 13 & 13:53/14:27 &  45$\degree$W & -1.34$\degree$ & 1.09$^{\prime\prime}$ & G430L/G750L & 0.1$^{\prime\prime}$ & 0.05$^{\prime\prime}$ & 500 & 7/6\\ \hline
\end{tabular}
\end{center}
\end{table*}

Flux- and wavelength-calibrated x2d files were provided by HST following reduction with the standard STIS calibration pipeline (calstis). We reprocessed the G750L data using the same pipeline, but executing additional defringing procedures, in order to remove substantial fringes from the longest wavelengths. We extracted individual spectra from each row of the two-dimensional spectral images, corresponding to the 0.025$^{\prime\prime}$ and 0.05$^{\prime\prime}$ pixel scales of the UV and visible gratings, respectively ($\sim$80 km and $\sim$160 km at the sub-observer point and comparable to the diffraction-limit at the grating central wavelengths). We then divided each extracted spectrum by a solar spectrum at the appropriate wavelengths to convert to reflectance. For the mid-UV G230L spectra, we use solar spectra for the corresponding date, as measured by the Total and Spectral Solar Irradiance Sensor \citep[TSIS-1;][]{RichardEtAl2020} on the International Space Station and obtained via the LASP Interactive Solar Irradiance Datacenter. Prior to division, we smoothed and binned the TSIS spectra to match the STIS spectral resolution and sampling. For the visible-wavelength G430L and G750L spectra, we used the ASTM E490 zero-airmass solar reference spectrum\footnote{\url{https://www.nrel.gov/grid/solar-resource/spectra-astm-e490.html}}. As the G430L and G750L data were taken using identical slit scans during the same visits, we combined the extracted spectra in these settings, scaling as necessary to correct small flux offsets, in order to produce complete 300--1000 nm spectra for each surface location. We calculated the geographic coordinates of each extracted pixel on Io using the aperture geometry information in the HST FITS headers and the phase, angular size, and pole orientation of Io obtained from JPL Horizons\footnote{\url{https://ssd.jpl.nasa.gov/horizons/}}.

Beyond 700 nm, the G750L spectra appeared to exhibit multiple artifacts leftover from the defringing process, similar to those also seen in STIS spectra of Mars and Europa \citep{Bell2007, Trumbo2020}. In addition, the broad point-spread function resulted in significant slit losses throughout the G750L wavelengths ($\sim$550--1000 nm) that resulted in artificial dropping reflectance toward the near infrared. To correct for these effects, we compared a constructed disk-integrated spectrum of Io's trailing hemisphere from the HST data to corresponding ground-based spectra. We fit a spline curve to the ground-based trailing-hemisphere spectrum of \citet{Spencer1995}, extrapolated the fit beyond 740 nm, which proved largely consistent with the somewhat sparse spectral sampling of \citet{ClarkMcCord1980} at these longer wavelengths, and then multiplied each HST spectrum by the ratio of the fitted spline to the corresponding full-disk spectrum we constructed from the HST observations. This approach simultaneously corrected for the wavelength-dependent slit losses and defringing artifacts, while maintaining all relative pixel-to-pixel variations and approximate spectral shapes within the G750L data. An additional source of artifacts complicated pixels very near to the limb of Io, where the signal was significantly reduced. Beyond $\sim$900 nm, these pixels appeared to contain dominant contributions from the so-called ``red halo" of scattered light within the STIS CCD, which created artificial spectral shapes. Though our reduction does not correct for this effect, it hinders only the usefulness of the most extreme limb data at the longest wavelengths.

\section{Definition and Geography of Spectral Units} \label{sec:units}

Coarse spectra produced from both the \textit{Voyager} and \textit{Galileo} imagery were previously used to define distinct spectral units on Io's surface. The \textit{Voyager} imagery could largely be described by smooth variation between three main endmembers---one interpreted as primarily SO$_2$ frost at low latitudes, a second of an unknown composition possibly rich in elemental sulfur and apparently associated with plume deposits, and a third of unknown composition constrained to the visibly darkened high latitudes \citep{McEwen1988color}. The \textit{Galileo} images identified four main color units: white materials, yellow materials, red materials, and dark materials, each of which displayed distinct characteristics in the SSI filters \citep{Geissler1999}. The white materials, distinguished by their relatively shallow slopes between the short-wavelength filters, closely matched the low-latitude \textit{Voyager} SO$_2$ unit and were confirmed as SO$_2$-rich based on infrared \textit{Galileo} spectra. The yellow materials, characterized by a steep short-wavelength slope and a sharp inflection at green wavelengths, were shown to dominate the low-latitudes outside of the SO$_2$ deposits. The red materials, featuring a steep slope from the violet to red filters, were found to dominate the high latitudes, Pele's ring, and smaller red deposits associated with volcanic centers. In contrast, the dark materials, accounting for just 1\% of Io's surface, exhibited a slight local minimum in the 889 nm SSI filter, which was interpreted as a possible Fe$^{2+}$ crystal field absorption of exposed orthopyroxenes \citep{Geissler1999, carlson2007io}.

\begin{figure*}[ht]
\figurenum{1}
\plotone{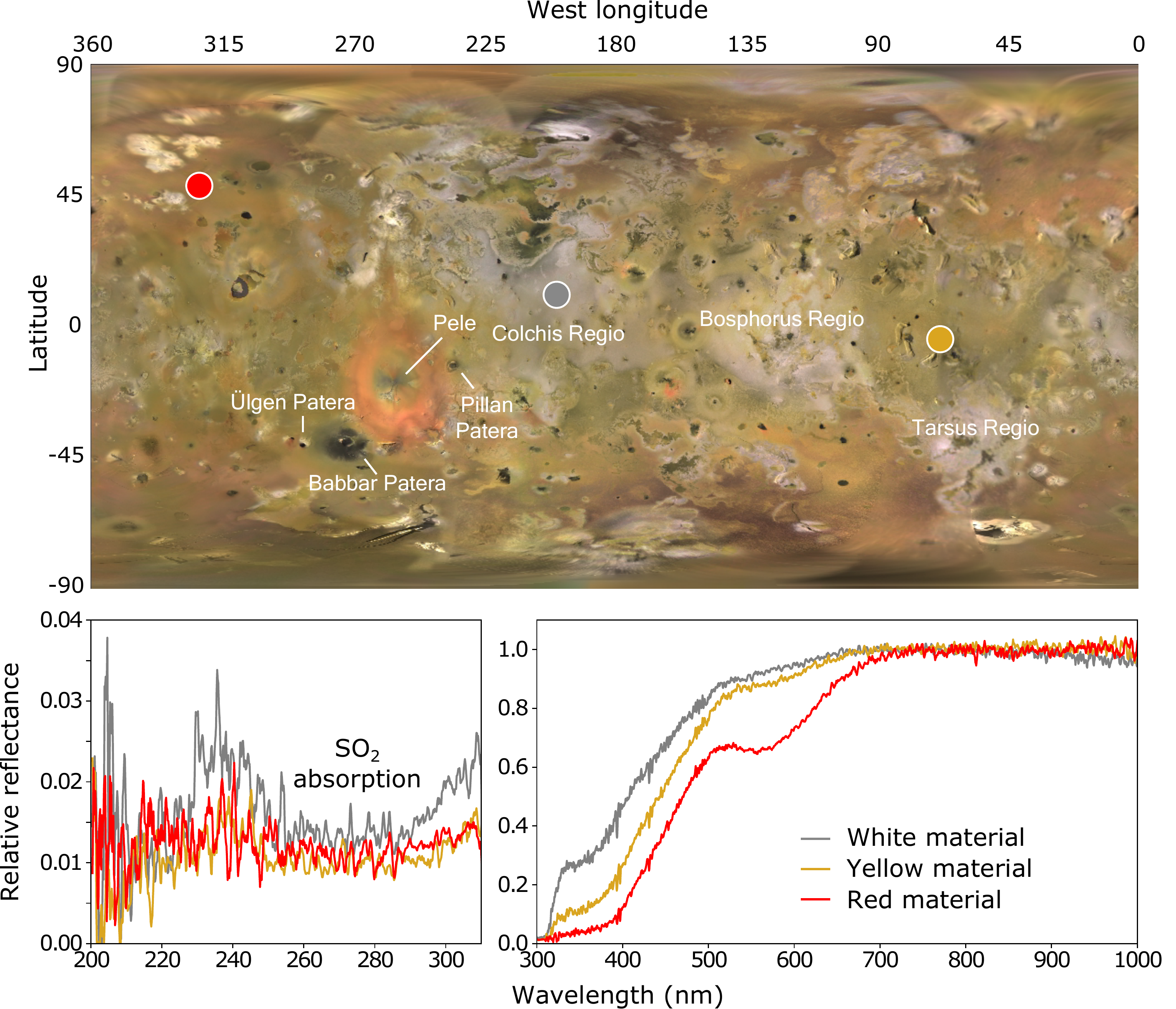}
\caption{Example HST spectra of the white, yellow, and red color units defined from \textit{Galileo} imagery \citep{Geissler1999}. Colored dots on the background USGS mosaic of Io indicate the pixel locations for the example spectra. The white materials show generally shallower slopes than do the red and yellow materials, with the exception of a steep drop short of $\sim$320 nm that is indicative of SO$_2$, as well as strong SO$_2$ absorption (low relative reflectivity) near 280 nm in the UV. The yellow materials are primarily characterized by a steep positive slope from roughly 380 to 500 nm, while the red materials exhibit a similarly strong slope in addition to a strong absorption near 560 nm. We performed 7-pixel and 5-pixel moving average smoothing to the shown UV and visible spectra, respectively. \label{fig:units}}
\end{figure*}

In Figure \ref{fig:units}, we show representative HST/STIS spectra of the white, yellow, and red color units defined and mapped previously by \cite{Geissler1999} from \textit{Galileo} imagery. As expected, the spectrum of the so-called white material, noted by \citet{Geissler1999} to realistically be more pale pink or yellow in appearance, exhibits strong SO$_2$ absorption in the UV, with characteristic reflectance minima near 220 and 280 nm and a local maximum near 240 nm \citep{Nash1980, Hapke1981}, and weaker spectral slopes than the rest of the surface in the 400–700 nm range. However, its spectral slopes still indicate some absorption in this visible wavelength range that is not expected for pure SO$_2$ frost. In addition, a very steep absorption edge shortward of 320 nm appears in the visible-wavelength spectra of the white materials and is also indicative of SO$_2$ frost, as this edge marks the onset of the aforementioned 280 nm UV SO$_2$ band \citep{Nash1980, HapkeEtAl1981}. The example spectrum of yellow materials is less characteristic of SO$_2$ frost and is instead dominated by the strong 380--500 nm absorption edge observed previously in ground-based spectra \citep[e.g.][]{Wamsteker1974, Spencer1995}. The other absorption visible in the ground-based data, described as either a 550--700 nm absorption edge \citep[e.g.][]{Spencer1995} or a 560 nm absorption feature \citep[e.g.][]{carlson2007io}, is only weakly visible in spectra of the yellow materials, but appears to be a strong and defining feature of the red materials seen by \textit{Galileo} (Figure \ref{fig:units}). These red regions also exhibit a strong 380--500 nm absorption edge, though it is interesting to note that the inflection at the bottom of the band edge appears to shift to longer wavelengths as one moves from white to yellow to red materials. Beyond 700 nm, the spectra of all three units are largely flat, which is consistent with both the ground-based and \textit{Galileo} data. 

Though the HST spectra span the wavelengths needed to confirm the tentative 890 nm absorption associated with the dark, likely silicate, materials, most such regions are on the order of tens of kilometers in size \citep[e.g][]{Geissler1999} and thus are below the spatial resolution of our data, which approaches the $\sim$300 km diffraction limit for 890 nm wavelengths at the sub-observer point. Thus, it is reasonable to expect the weak 890 nm minima observed by \textit{Galileo} to be diluted beyond detection in our spectra, and, indeed, we see no clear spectral evidence for silicates. The only location that may present a possible exception is Babbar Patera (38\degree S, 277\degree W), whose 400-kilometer-wide diffuse dark deposits are on par with the angular resolution of our data \citep{McEwen1998}. Unfortunately, however, the longitudes of Babbar appear near the limb of Io as observed from Earth for the dates and times of our observations, such that the increasing contributions of the long-wavelength scattered light described in Section \ref{sec:observations} make the possible appearance of an absorption there too uncertain to constitute a detection. Future observations at more targeted geometries may be able to better address the outstanding question of the possible 890 nm silicate absorption in dark materials.
\newline
\subsection{Endmember Extraction}

In addition to examining the spectral characteristics of previously defined \textit{Galileo} color units, we take our own approach to identifying spectral endmembers in the HST data. As the UV spectra provide lower spatial resolution and appear to be dominated simply by variations in SO$_2$ absorption, we focus solely on the visible wavelengths for our endmember analysis. We follow a similar method to that described in \citet{Fischer2015}, in which we decompose the spectra at each spatial pixel into a linear combination of the N most extreme spatial pixels observed. To identify the N optimal endmember spectra, we first exclude any pixels observed at greater than 60$^{\circ}$ from the center of Io’s disk. We then take an iterative approach for identifying the optimal set of endmember pixels. We input all of the spectra into a K-means clustering algorithm and find N wavelength-dependent cluster centers. We then identify the spatial pixels whose spectra are most similar to each cluster center using the spectral angle mapping (SAM) criteria first described in \citet{Kruse1993}. The N spatial pixels with the smallest SAM angle $\theta$ from each k-means cluster center are then used as the initial set of spectral endmembers and the remaining spatial pixels are modeled as a linear combination of these endmembers using standard least-squares regression. The residual root-mean-square (RMS) error is calculated for each pixel, and the RMS of all pixels is summed to define a total RMS for the given set of endmembers. 

\begin{figure}[t]
\figurenum{2}
\plotone{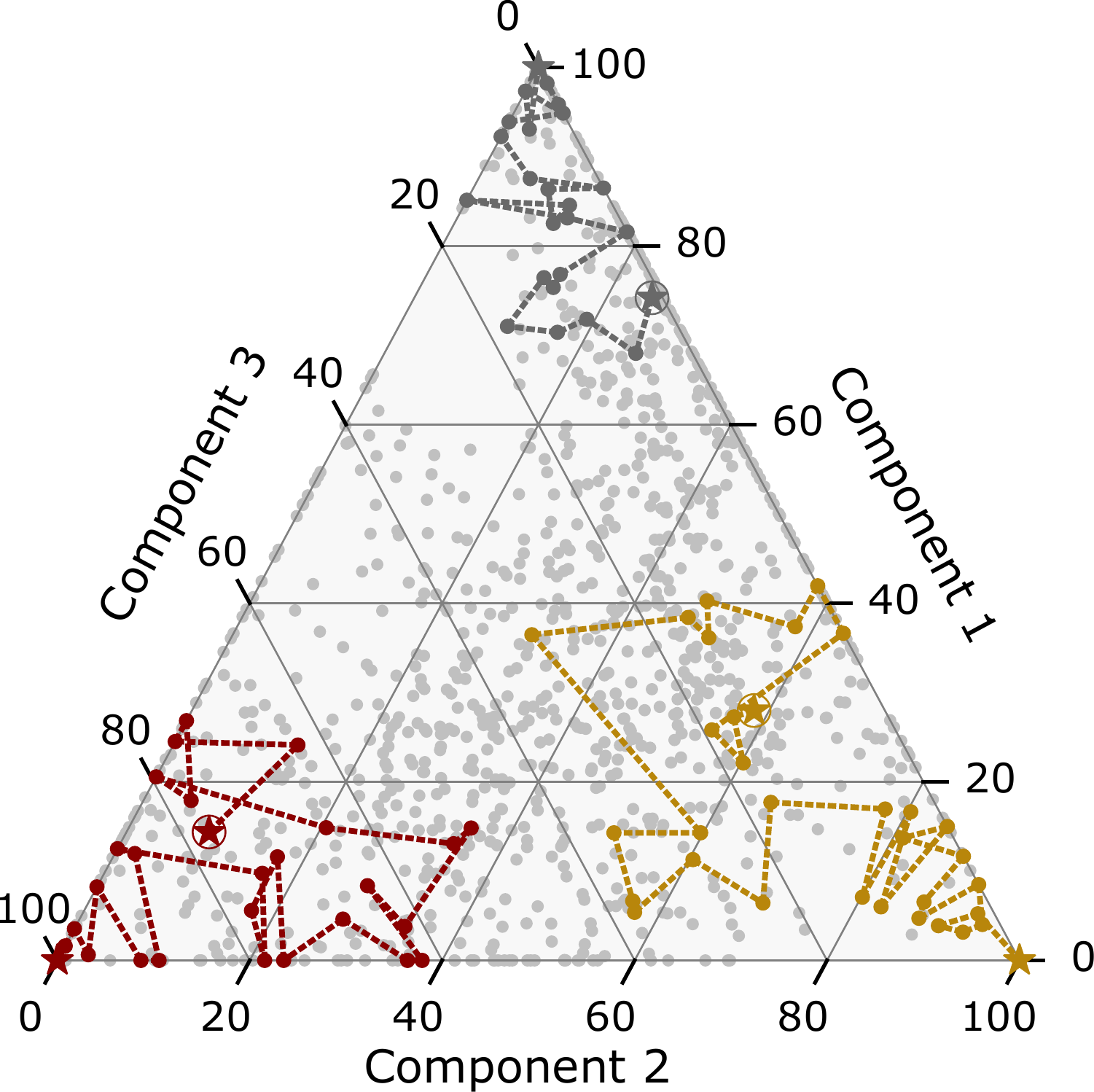}
\caption{Ternary diagram demonstrating the progression of our endmember extraction routine, in which the endmembers converge to the corners of the diagram. The initial set of spectral endmembers is indicated by the encircled stars, while colored dots show iteratively chosen subsequent endmembers that lowered the overall RMS. Grey dots represent the remaining pixels as best-fit linear combinations of the final three components. \label{fig:ternary}}
\end{figure}

This initial endmember set is then iteratively improved by holding all but one endmember constant, and randomly selecting a new endmember candidate from the set of pixels composed of at least 70\% that endmember in the linear models. A candidate endmember is accepted if the total RMS error using the new candidate set is smaller than that of the previous set of endmembers. This process continues, alternating which endmember is substituted, until no new endmember candidates are accepted for $100 \cdot N$ iterations and the total deviation in RMS between the last few accepted iterations has leveled off below $\sim$1\%. Figure \ref{fig:ternary} illustrates this minimization procedure for the $N=3$ case, where the optimal endmember candidates converge to the 3 corners of the ternary diagram.

To ensure convergence to the optimal endmember set, we run the minimization procedure 100 times and find that 92 of the test cases converge to the lowest total RMS. The remaining 6 test cases find an endmember set that is very similar, both qualitatively and quantitatively, to the lowest RMS set, but which has a slightly higher total RMS value. We also perform this spectral decomposition routine for endmember sets with values of N ranging from 1 to 6. The total RMS error of the final endmember set drops significantly from the N=1 to N=2 case, and again from the N=2 to N=3 case. The residual maps also reveal significant spatial structure for the N=1 and N=2 cases, revealing the likely presence of additional endmembers. When moving from the N=3 to N=4 and 5 cases, the improvement in the total RMS is small. While there is still some spatial structure in the residual map for the N=3 case, we cannot confidently determine whether this is due to the presence of a fourth endmember or an artifact of systematic noise, so we choose N=3 endmembers for our final spectral decomposition model.

\subsection{Endmember Spectra and Distributions}

\begin{figure}[t]
\figurenum{3}
\plotone{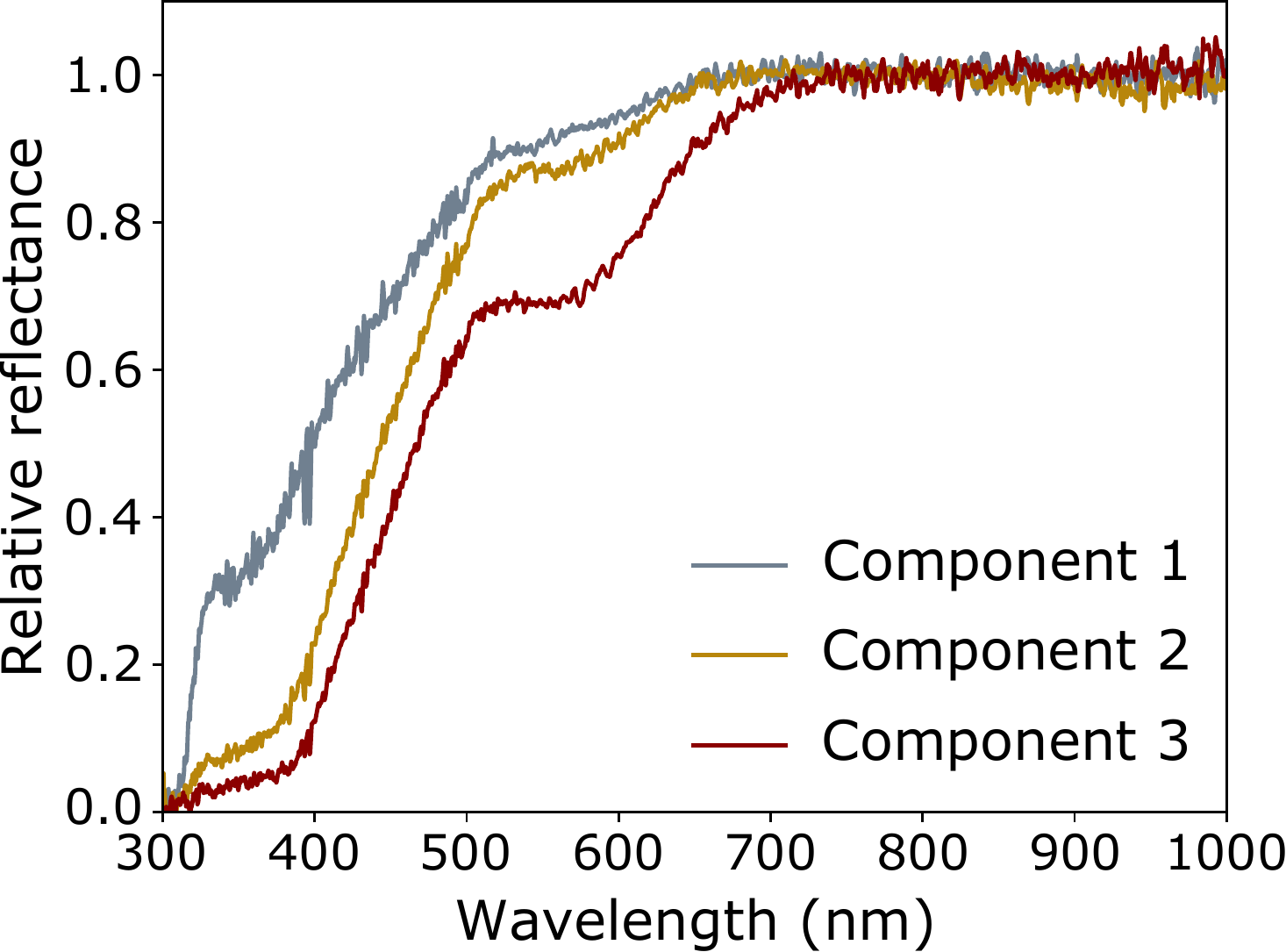}
\caption{The final three endmembers extracted from the visible-wavelength HST dataset. The spectra have been smoothed by a 5-pixel moving average filter. The first, second, and third components appear similar to spectra of the \textit{Galileo} white, yellow, and red materials, respectively. \label{fig:endmembers}}
\end{figure}

\begin{figure}[t]
\figurenum{4}
\plotone{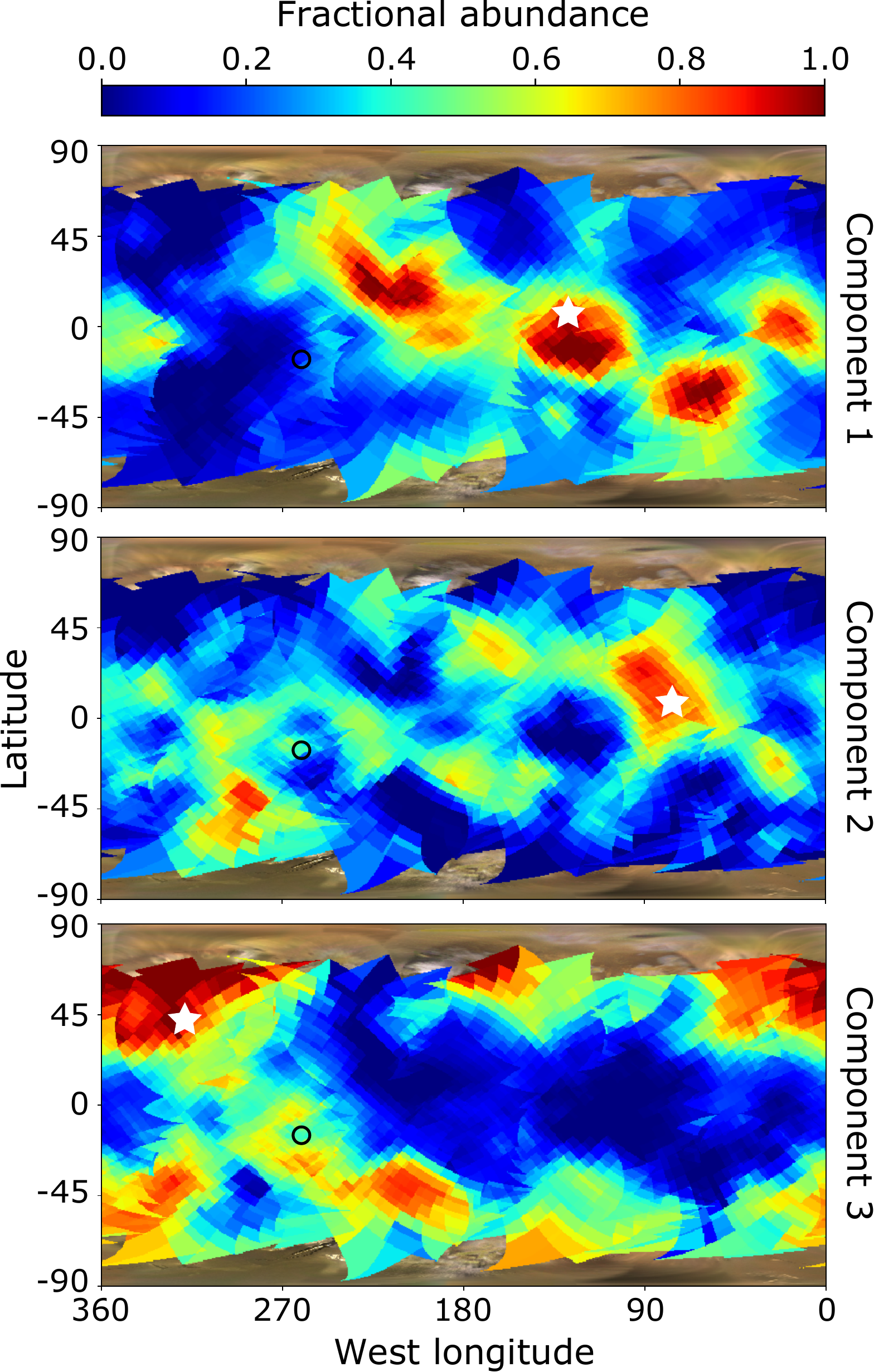}
\caption{The geographic distributions of our extracted spectral endmembers. The given fractional abundances are the coefficients of the best-fit linear combinations of the three components at each pixel. White stars indicate the location of the endmember pixels, and a black circle indicates the location of Pele (19$\degree$S, 255$\degree$W). Component 1 closely follows the \textit{Voyager} SO$_2$ unit identified by \citet{McEwen1988so2} and also matches our SO$_2$ map in Figure \ref{fig:so2} well. Component 2 correlates with the yellow, likely sulfur-rich plains immediately surrounding the SO$_2$ deposits, while our map of component 3 highlights the upper latitudes and Pele's pyroclastic ring (19$\degree$S, 255$\degree$W) in good correspondence to our map of Io's 560 nm absorption in Figure \ref{fig:560}. \label{fig:endmember_maps}}
\end{figure}

The endmember approach described above arrived at the three spectrally distinct endmember components shown in Figure \ref{fig:endmembers}. In essentially every respect, they are quite similar to the three spectra displayed in Figure \ref{fig:units}---component 1 displays shallower slopes than the others, except for a steep drop near 320 nm (akin to the white materials), component 2 exhibits a dramatic 380--500 nm absorption edge (similar to the yellow materials), and component 3 possesses the strongest 560 nm absorption (like the red materials). Maps of the fractional abundance of each endmember in linear models of each pixel reveal geographic distributions consistent with the primary units defined by \citet{Geissler1999} (Figure \ref{fig:endmember_maps}). As with the white materials defined from \textit{Galileo} imagery, component 1 largely follows the \textit{Voyager} SO$_2$ unit \citep{McEwen1988so2}, while component 2 correlates with the yellow regions seen to surround such SO$_2$-rich areas, largely avoiding the highest latitudes dominated by both the red materials and component 3 from our endmember analysis (Figure \ref{fig:endmember_maps}). In fact, component 2 appears to nearly always flank the fields dominated by component 1, with a slight exception north of the equatorial patch at $\sim$15$\degree$W.

Unsurprisingly given its red color and similarity to the upper latitudes in coarse \textit{Galileo} imaging spectra, Pele's pyroclastic ring at roughly 19$\degree$S and 255$\degree$W also appears distinct in our map of component 3. It exhibits the strongest association with component 3 in its southernmost portion and appears possibly broken in the northeast near Pillan Patera (12$\degree$S, 243$\degree$W), which was seen to disrupt Pele's ring following its 1997 eruption during the \textit{Galileo} mission \citep[e.g.][]{Davies2001, Keszthelyi2001, Williams2001} and which has subsequently been seen to erupt episodically from ground-based observations \citep{dePater2016}. In contrast, Pele's center and the regions along the outer edge of its ring deposit appear enriched in component 2. The most obvious exception to the correspondence between our endmember maps and the color unit maps of \citet{Geissler1999} occurs just southwest of Pele's ring deposit and Babbar Patera (38\degree S, 277\degree W) near Ülgen Patera (41\degree S, 287\degree W), which appears largely red in the \citet{Geissler1999} maps, but is dominated by component 2 in ours. This discrepancy may imply modification to Io's surface composition at this location in the decades between the \textit{Galileo} and HST observations. Elsewhere, and with the exception of Pele's ring and the northwestern portion of Colchis Regio, which appears rich in component 1, component 3 nearly always flanks component 2 on the pole-ward side. Indeed, the combination of our three component maps agrees well with the color unit geographies from \textit{Galileo}, suggesting that the surface compositional variation is well-captured by these three primary units. In addition, they bear resemblance to some of the large-scale geologic units mapped on Io \citep{Williams2011geologic, Williams2021io}, with component 1 encompassing much of the mapped ``white bright plains material" in addition to some regions classified as ``bright" or ``undivided" flow material, component 2 exhibiting fair correspondence to the ``yellow bright plains material", and component 3 appearing largely in regions characterized as ``red-brown plains" (with the key exception of Pele's ring).\newline
\newline

\section{Spectral Feature Mapping} \label{sec:features}

In addition to mapping spectral endmembers, which are determined based on the entire 300--1000 nm continuum, we also map the strengths of individual spectral features across the surface. Despite providing the first global, spatially resolved spectra at these wavelengths, the HST data do not reveal any discrete absorptions not already seen in past Earth-based observations \citep[e.g.][]{Nelson1980,Spencer1995}. The mid-UV absorption signatures and 320 nm absorption edge of SO$_2$, the 380--500 nm absorption edge, and the 560 nm feature are the only features apparent in our data.

We map the distribution of Io's SO$_2$ frost using the contrast of its strong absorption maximum near 280 nm to the surrounding continuum. Some of the fine structure visible below 240 nm in our UV spectra may be sensitive to band systems of Io's atmospheric SO$_2$ \citep[e.g.][]{Ballester1994so2}, but the measurement of this is outside the scope of this study of surface composition. Though atmospheric SO$_2$ also exhibits broad absorption across the mid-UV, in addition to sharp structure, surface reflectance has been shown to dominate the strength of the wide $\sim$280 nm band \citep[e.g.][]{Jessup2004, JessupSpencer2015}.

We fit a linear continuum from 236 to 315.5 nm, excluding the interval corresponding to the absorption (239--313.5 nm), remove this continuum, and integrate the residual absorption to obtain the equivalent width (width of a 100\% absorption feature of the same area). We note that, as this band is the culmination of the overall dropping reflectance across the near- to mid-UV that begins at 320 nm, this continuum is likely non-physical, but simply serves as a tool against which to measure the relative relief of the absorption band.

We also map the strength of the two visible-wavelength absorptions, the 380–500 nm edge and the 560 nm feature, which are not characteristic of SO$_2$ frost. Given the uncertainty in defining the unknown missing continua associated with these features, we measure each in two different ways in order to ensure a robust assessment of their geographies. For the 380–500 absorption edge, we first define and remove a linear continuum, as above, from 324.5 to 440 nm, excluding the 332.5--425 nm region visibly associated with the inflection of the absorption, and measure the residual area. Second, we normalize the spectra to the mean reflectance between 488 and 490 nm near the top of the absorption and measure the resulting slope from 400 to 490 nm, interpreting steeper slopes as stronger absorption. Slopes across slightly different wavelength bounds within the band edge produce comparable results. Similarly, for the 560 nm feature, we assess the equivalent width by fitting and removing a third-order polynomial from 460 to 745 nm, excluding the visible absorption across 500--700 nm, and also normalize the spectra at the top of the feature (663--665 nm) and measure the slope from 585 nm within the absorption to 665 nm at the top. Second-order polynomial continua result in equivalent, though slightly noisier, results. We then map all of these measurements using the coordinates as obtained in Section \ref{sec:observations} and compare their distributions with those of each other, past observations, geology, and likely surface processes. 

\subsection{SO$_2$ Frost}
\begin{figure}[t]
\figurenum{5}
\plotone{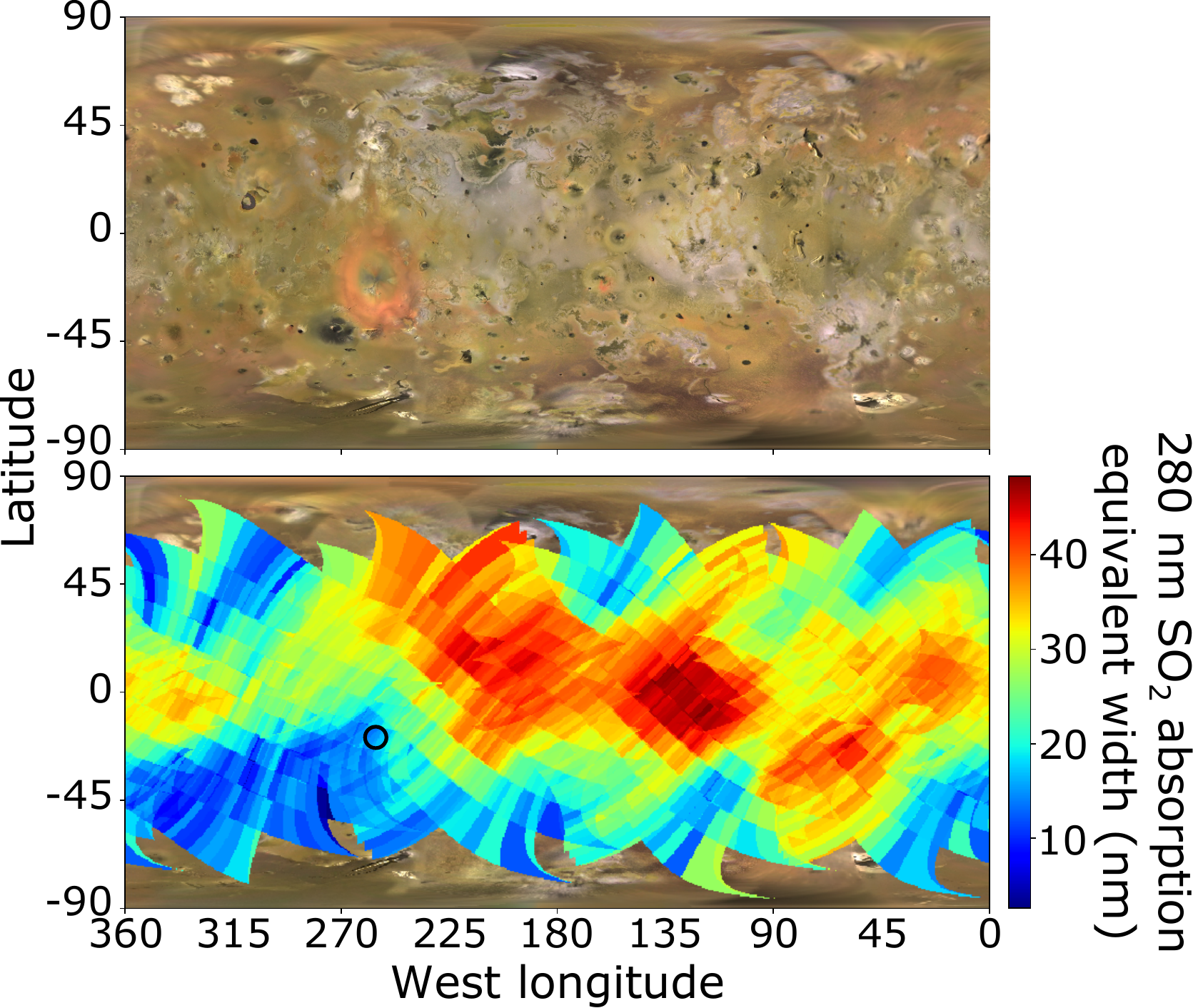}
\caption{Map of Io's 280 nm SO$_2$ absorption maximum from the UV HST/STIS G230L spectra. This distribution corresponds extremely well to the \textit{Voyager} SO$_2$ frost map of \citet{McEwen1988so2} and to the visibly bright regions of the above USGS \textit{Galileo/Voyager} mosaic. We observe strong SO$_2$ bands within Colchis Regio and Bosphorus Regio at low-latitudes between roughly 100$\degree$ and 250$\degree$W, which is also in agreement with ground-based mapping of weak infrared SO$_2$ bands \citep{Laver2008, Laver2009, dePater2020}. Such agreement, in combination with the relative strengths of different SO$_2$ bands observed by \textit{Galileo} NIMS \citep{Carlson1997Io,Doute2001}, is consistent with these regions harboring the largest-grained, thickest, and/or purest SO$_2$ deposits, which appear stable over multi-decade timescales. For context, the location of Pele is indicated by a black circle. \label{fig:so2}}
\end{figure}

Figure \ref{fig:so2} shows the SO$_2$ distribution we obtain based on the 280 nm mid-UV feature. In agreement with distributions derived from \textit{Voyager} and HST imaging \citep{McEwen1988so2, Sartoretti1996} and \textit{Galileo} and ground-based mapping of Io's infrared SO$_2$ bands \citep{Carlson1997Io,Laver2008, Laver2009, dePater2020}, we observe strong SO$_2$ absorption at equatorial latitudes within Bosphorus Regio and Colchis Regio between $\sim$100 and 250$\degree$W. In fact, our SO$_2$ map shows a distribution qualitatively identical at our resolution to that produced from the Voyager imagery \citep{McEwen1988so2} and to the visibly bright regions of the USGS \textit{Galileo}/\textit{Voyager} color mosaic of Io\footnote{\url{https://astrogeology.usgs.gov/search/map/Io/Voyager-Galileo/Io\_GalileoSSI-Voyager\_Global\_Mosaic\_ClrMerge\_1km}}, with all three showing additional deposits near the sub-Jovian point, at localized high northern latitudes, and within Tarsus Regio ($\sim$40$\degree$S, 55$\degree$W), which are less apparent in the infrared maps. In addition, the SO$_2$ map we produce corresponds extremely well to the geography of the first spectral endmember (component 1) from Section \ref{sec:units}. Based on comparisons between the derived surface reflectance and atmospheric transmission signatures for the SO$_2$-rich Prometheus plume shown in \citet{Jessup2004} (their Figure 4), we estimate that unaccounted for atmospheric contributions may introduce an upper limit of $\sim$15\% overestimation in our absolute band strengths. However, the near identical correspondence to the aforementioned additional proxies for SO$_2$ frost mapped previously suggests that atmospheric uncertainties are a small factor in determining the relative geography we obtain.

It is interesting to note, however, that the distribution we observe lacks resemblance to that produced via inversion of \textit{Galileo} NIMS data with linear spectral models, rather than by mapping of individual band strengths \citep{Doute2001}. In their retrievals, \citet{Doute2001} find high fractional SO$_2$ coverage at mid latitudes to high latitudes and relative depletions at the more equatorial Colchis Regio and Bosphorus Regio, where we find the strongest 280 nm bands. From their models, the authors also estimate larger equivalent widths at the equator and much larger equivalent widths southeast of Pele than those measured by \citet{Carlson1997Io} from comparable NIMS spectra. To explain such differences, the authors suggest a combination of an opposition surge decreasing the band strengths observed by \citet{Carlson1997Io} and an optically thin, yet often large-grained, frost at mid to high latitudes that is insufficient to result in some of the weaker infrared SO$_2$ bands measured by \citet{Carlson1997Io}, but that they suggest can be deduced from the modeling of stronger bands. To justify the differences between their retrieved distribution and those from the \textit{Voyager} and \textit{Galileo} imaging results, \citet{Doute2001} invoke a different explanation, instead proposing that molecular contamination of the mid- and high-latitude SO$_2$ deposits with darker sulfur suppresses their near-UV and visible albedos such that they do not appear in maps produced from images at those wavelengths.

It is not clear, however, that either explanation adequately accounts for the discrepancy with the map we produce from the strong mid-UV SO$_2$ band. At such short wavelengths, our observations should be sensitive to frost deposits, even if they are optically thin at the infrared wavelengths observed with \textit{Galileo} NIMS. In addition, though sulfur is darker than SO$_2$ in the visible, SO$_2$ frost is comparably dark in the mid-UV \citep[e.g.][]{Nash1980}, which calls into question the idea that molecular mixing with sulfur may be responsible for masking the otherwise high near-UV and visible reflectivity of SO$_2$ frost at the mid to high latitudes in \textit{Voyager} data. Thus, our results are possibly contradictory to those of \citet{Doute2001} and may imply additional effects not yet considered or perhaps limitations to the retrieval technique they employed. However, our band-strength mapping cannot distinguish between the often degenerate effects of abundance and grain size, and we do not attempt to constrain these effects with our limited data, which undoubtedly also contain spectral contributions from other surface species that may be affecting the UV reflectance at the upper latitudes. Indeed, the apparent SO$_2$ concentrations within Bosphorus Regio and Colchis Regio in our map do correspond well to the distributions of the largest equatorial to mid-latitude SO$_2$ grains inferred by \citet{Doute2001}. In addition, the UV spectral shape of SO$_2$ frost varies slightly between different laboratory experiments \citep[e.g][]{Nash1980, Hapke1981}, depending on the frost properties, and Io's geographically varying surface conditions have not been fully replicated in laboratory settings.

Future work involving simultaneous modeling of the UV, visible, and infrared spectra and the consideration of grain size and frost structure, which is outside the scope of this paper, may resolve the apparent discrepancies in Io's SO$_2$ frost distribution. Nevertheless, all of the observations, particularly the relative strength of the weaker infrared SO$_2$ bands in the ground-based and NIMS data, are broadly consistent with the thickest and/or largest-grained and possibly purest SO$_2$ deposits at equatorial latitudes. Furthermore, the remarkable agreement between the \textit{Voyager}, \textit{Galileo}, and HST maps indicates long-term stability of frost in these regions, despite the constant activity and high resurfacing rates on Io.  

\subsection{380--500 nm Absorption Edge}

\begin{figure}
\figurenum{6}
\plotone{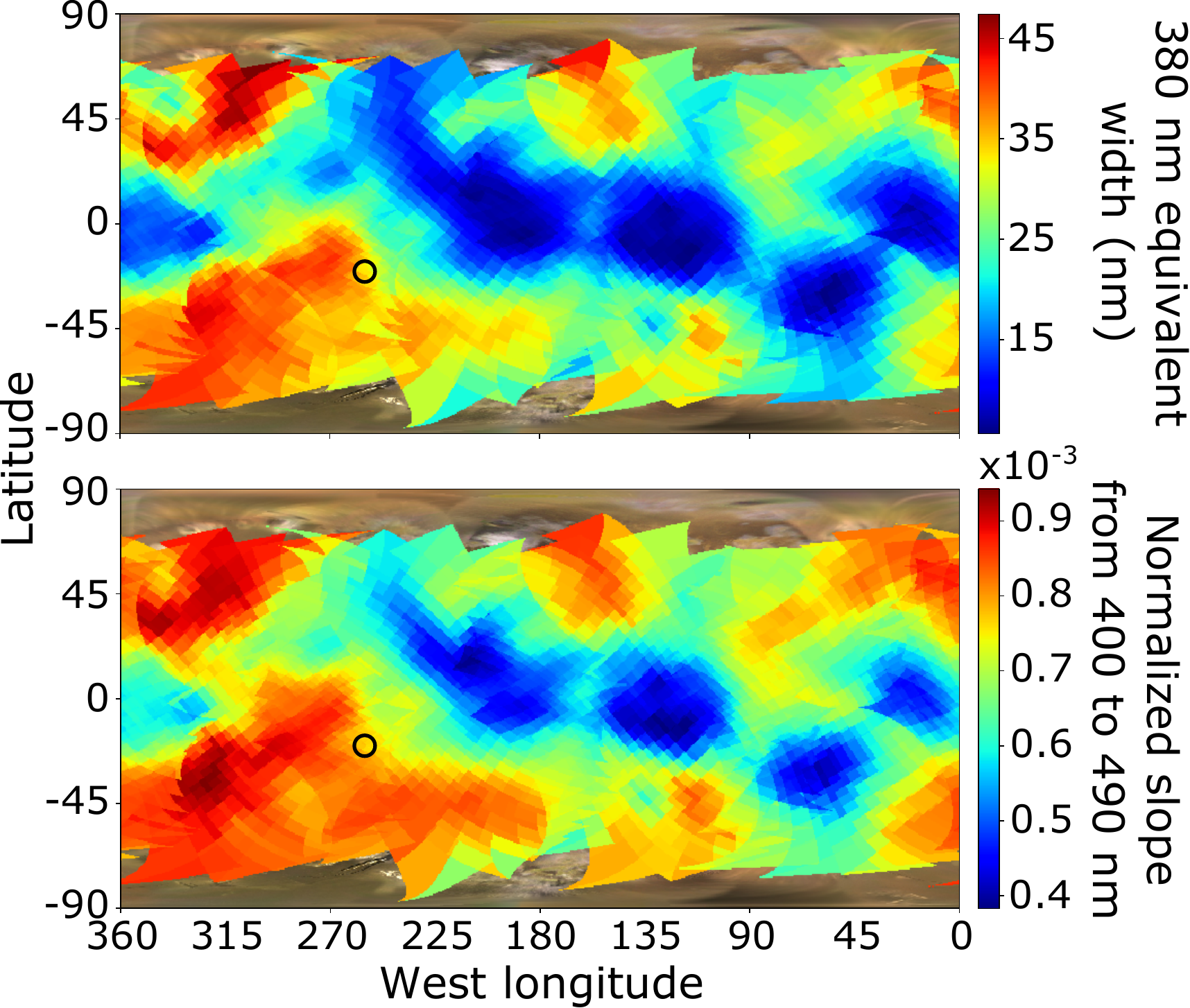}
\caption{Maps of Io's 380--500 nm absorption edge, which we measure in two ways. We create the top map by measuring the band area beneath a fitted linear continuum across the inflection of the absorption, while we create the bottom map by measuring the slope of the band edge normalized at its long-wavelength end. The distributions of both proxies agree extremely well, demonstrating that the absorption is entirely anti-correlated with the SO$_2$ signature mapped in Figure \ref{fig:so2}. Instead, it appears strong across much of the mid and high latitudes on Io and in some low-latitude locations not covered by SO$_2$ deposits. This geography is consistent with a wide distribution of sulfurous material, potentially rich in S$_8$ or S$\mu$. The location of Pele (19$\degree$S, 255$\degree$W) is indicated by the black circle. \label{fig:380}}
\end{figure}

Maps of our two proxies for the strength of the 380--500 nm absorption edge are shown in Figure \ref{fig:380}. Both show qualitatively similar distributions with minima associated with the aforementioned SO$_2$ deposits and enhanced absorption at some mid- to high-latitude locations, within the trailing/sub-Jovian quadrant, and southeast of Pele (19$\degree$S, 255$\degree$W). Our maps are broadly consistent with the 0.410/0.564 $\micron$ ratio map produced with HST images of Io \citep{Spencer1997} and with the understanding that overlying deposits of bright SO$_2$ mask underlying darker materials \citep[e.g.][]{carlson2007io}. The absorption edge appears strongly in both regions associated with the yellow materials and in those containing the red materials defined from \textit{Galileo} imagery \citep{Geissler1999}. Indeed, simply mapping the strength of this feature does a poorer job of spatially distinguishing between these two units than does our endmember approach of Section \ref{sec:units}, which maps yellow materials as having a higher relative proportion of the endmember defined almost exclusively by its strong 380--500 nm edge and finds the red materials to have a higher relative proportion of the endmember defined by a combination of a strong edge and 560 nm feature. This difference underscores how both methods---endmember classification and spectral feature mapping---can complement each other in the process of assessing geographic compositional variations.

As noted in Section \ref{sec:units}, the inflection at the bottom of the 380--500 nm absorption appears to shift in wavelength between roughly 360 and 390 nm, with shorter-wavelength minima corresponding to the SO$_2$ deposits and the longest-wavelength minima generally corresponding to regions of strongest absorption, particularly within the trailing/sub-Jovian quadrant. It is possible that the weak 360 nm feature of SO$_2$ frost \citep{HapkeEtAl1981}, previously unobserved on Io, is influencing the spectrum at that wavelength within the strongest SO$_2$ deposits. Indeed, some such spectra appear to show a slight slope break just short of 400 nm, possibly indicating the influence of a separate feature from the overall band edge (see spectrum of white materials in Figure \ref{fig:units}). Otherwise, the distributions we obtain for the 380--500 nm absorption edge, which are strongly anti-correlated with SO$_2$ and span much of the yellow and red-brown geologic plains units \citep[individually well-matched by components 2 and 3 above, respectively; ][]{Williams2011geologic, Williams2021io}, appear reasonable for the prior interpretations of this feature as characteristic of some form of elemental sulfur. It seems likely that the presence of such material is more or less ubiquitous on Io at our spatial resolution, given that even the apparently SO$_2$-rich white materials display the 380–500 nm band edge to some degree, but that it simply becomes partially obscured by overlying frost in deposits like those seen in the Bosphorus, Colchis, and Tarsus Regiones.
\newline

\subsection{560 nm Feature}
\begin{figure}
\figurenum{7}
\plotone{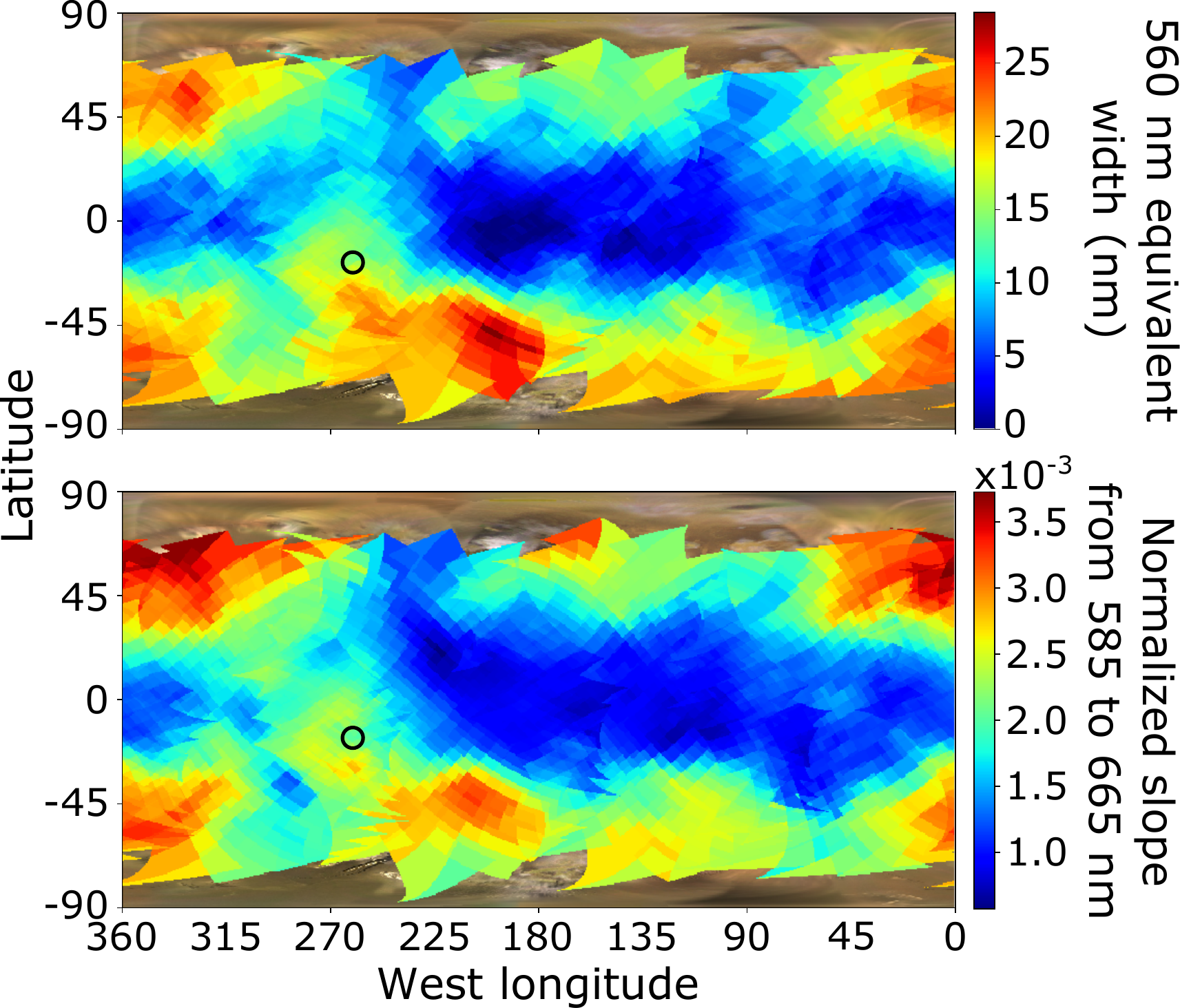}
\caption{Maps of Io's 560 nm absorption, which we measure in two ways. We retrieve the top distribution by measuring the band area beneath a fitted polynomial continuum across the visible absorption, while we retrieve the bottom distribution by measuring the slope of the 585--665 nm band edge normalized at its long-wavelength end. Both maps agree extremely well and indicate that the absorption is primarily constrained to the high latitudes with the notable exception of a clear association with Pele's ring of pyroclastic material, the approximate center of which is indicated by the black circle at 19$\degree$S and 255$\degree$W. This geography is potentially indicative of a thermodynamically unstable species that is continuously produced. In agreement with the interpretations of \citet{Spencer1997} and \citet{carlson2007io}, we suggest that the metastable short-chain allotrope S$_4$ may form directly from plume gases at Pele and possibly be independently produced by irradiation and/or preserved by low temperatures at the high latitudes (see Section \ref{sec:composition}). \label{fig:560}}
\end{figure}

Both of our methods for measuring the strength of Io's 560 nm absorption produced nearly identical distributions, which are shown in Figure \ref{fig:560}. Most strikingly, the high latitudes stand out as regions of strong absorption in our maps, while the low latitudes are almost entirely devoid of strong 560 nm features. The major exception, however, is the clear association of the 560 nm band with Pele's ring deposit (roughly 19$\degree$S, 255$\degree$W), which is just visible in both maps. In both respects, our maps largely agree with the 0.564/0.781 $\micron$ ratio map produced from the past HST imaging observations of \citet{Spencer1997}, though Pele's signature is somewhat weaker relative to the high latitudes in our data. This may reflect the decrease in Pele's activity implied by measurements of its thermal emission after 2002, which could have caused fading of its ring \citep{dePater2016}. In addition, we find stronger 560 nm absorption southeast of Pele's ring than southwest of it, which is opposite of the pattern in the HST imaging ratio map \citep{Spencer1997}. Though the spatial correspondence is not exact, it is also interesting to note that, despite Pele decreasing in activity itself, a region southeast of Pele (near 35$\degree$S, 240$\degree$W) appears to have initiated volcanic activity since the \textit{Galileo} mission \citep{deKleer2019io}. Convolution of our spectra with the WFPC2 filters used in the imagery shows that such differences are not explained by the wide imaging band passes and may instead imply such temporal changes in Io's surface in the more than 20 years between the imagery and our spectroscopic observations. Indeed, as mentioned in Section \ref{sec:units}, the region southwest of Pele near Ülgen Patera also appears discrepant between our endmember maps and those of previously defined \textit{Galileo} color units \citep{Geissler1999}. Regardless, however, the qualitative conclusions of the \textit{Galileo} and both HST datasets are consistent---Io's 560 nm absorption relates to fresh plume deposits at Pele and processes at high latitudes, and contributes to the reddish colors of such regions.\newline
\newline

\section{Implications for Composition and Surface Processes} \label{sec:composition}

The UV/VIS HST spectra of Io's surface presented here provide an unprecedented combination of spatial and spectral detail that can help constrain the volcanic species behind Io's colorful surface, as well as their relationships with surface processes. However, with the exception of the diagnostic SO$_2$ bands observed in the UV, the broad, non-unique, and potentially overlapping nature of Io's spectral features beyond 300 nm and those of candidate laboratory materials make concrete identification on the sole basis of spectral comparison essentially impossible. Instead, a combination of spectral plausibility from laboratory comparison and spatial correspondence to a conceptual model based in current understanding of Io surface processes presents our best means of unraveling the surface composition. As our UV spectra appear dominated by the simple modulation of the SO$_2$ signatures, and as many potentially relevant additional species lack laboratory spectra at UV wavelengths, we focus on our visible-wavelength data for such comparisons.

\begin{figure}
\figurenum{8}
\plotone{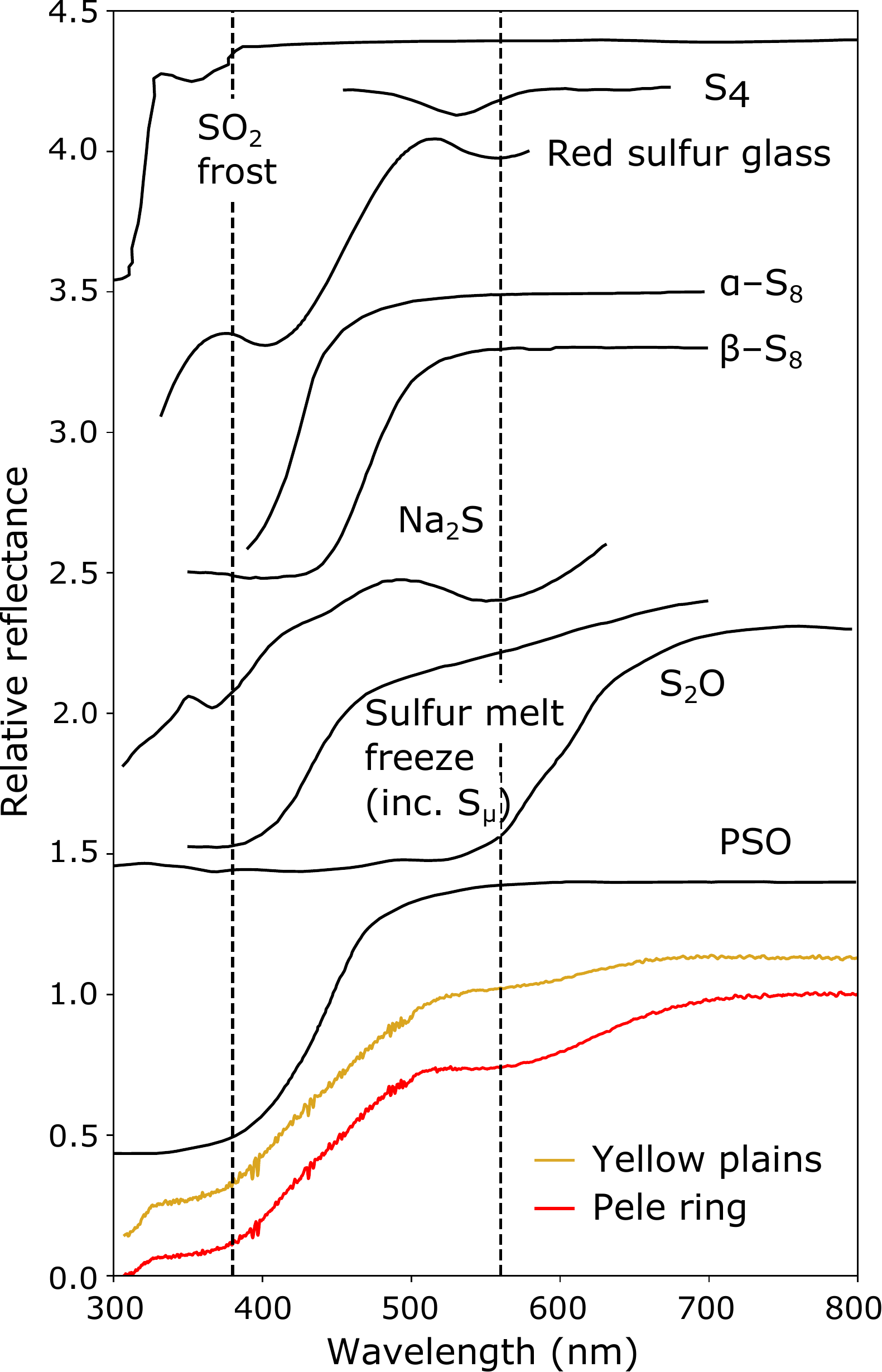}
\caption{Example average HST spectra of Io compared to laboratory spectra of candidate materials. Several species exhibit similar features to Io's 380--500 nm absorption edge, which has long been attributed to some form of sulfur. The inflection at the base of the band edge may be better matched by the presence of metastable polymeric sulfur (S$_\mu$; a significant component of the shown sulfur melt freeze) than by S$_8$ alone. Though the 560 nm band is perhaps best matched by the spectrum of Na$_2$S, considerations of Io's plume chemistry and of the geography of the feature point to the spectral influence of metastable S$_4$ (also seen in the spectrum of red sulfur glass and likely that of S$_2$O) as a more likely cause. The S$_4$ spectrum is from \citet{Carlson2009} after \citet{Meyer1972visible} and is scaled for clarity. The red sulfur glass spectrum is from \citet{Meyer1972}. Both reflect samples at 77 K and are converted to approximate reflectance from absorption profiles. The $\alpha$-S$_8$ spectrum is from \citet{Carlson2009} after \citet{Gradie1980} and was measured at room temperature, the spectra of $\beta$-S$_8$ and the sulfur melt freeze (from 453 K melt) are of room-temperature samples from \citet{MosesNash1991}, and the Na$_2$S spectrum shows an irradiated sample at 170 K from \citet{NashNelson1979}. The S$_2$O and PSO spectra are from \citet{HapkeGraham1989} and were taken at 77 K and 195 K, respectively. The shown SO$_2$ frost spectrum from \citet{HapkeEtAl1981} (at $\sim$77 K) lacks the above features. All spectra are normalized at their longest wavelengths and, with the exception of the Pele spectrum, are offset for clarity. The offsets used are 0.13, 0.4, 1.3, 1.4, 1.6, 2.3, 2.5, 3, 3.2, and 3.4 units, respectively, from bottom (yellow plains) to top (SO$_2$). Dashed lines at 380 and 560 nm pass through Io's primary visible absorptions. \label{fig:lab}}
\end{figure}

Io's 380--500 nm absorption edge has been widely accepted as characteristic of sulfur, though the exact form has been variably suggested to be stable elemental sulfur (orthorhombic cyclo-octal sulfur; $\alpha$-S$_8$), the more metastable allotropes monoclinic $\beta$-S$_8$ and polymeric sulfur (S$_\mu$), or some combination of these species, all of which possess a strong band edge roughly coincident in wavelength \citep[e.g.][]{Wamsteker1973, MosesNash1991,Spencer1997,carlson2007io, Baklouti2008}. Unfortunately, as the wavelength of the sulfur band edge changes with temperature, impurity, and irradiation, as well as with specific allotropes \citep{HapkeGraham1989, Kargel1999io,carlson2007io,Eckert2003}, its exact location cannot readily distinguish between such possibilities. $\alpha$-S$_8$ is expected to be the most thermodynamically stable form at Io surface conditions \citep{carlson2007io, Eckert2003}, but $\beta$-S$_8$ and S$_\mu$ rapidly cooled from high-temperature sulfur may still be long-lived on Io \citep{MosesNash1991}. In fact, as noted by \citet{MosesNash1991} and \citet{Baklouti2008}, S$_\mu$ may provide a better explanation for the 330--400 nm region of Io's spectrum, which exhibits a significantly less abrupt slope transition than do spectra of S$_8$ (Figure \ref{fig:lab}). Alternatively, \citet{Hapke1989} suggested polysulfur oxide (PSO) formed via dissociation of SO$_2$ in place of elemental sulfur, in part to explain the 330--400 nm region. However, widespread PSO in such high abundance as to explain the ubiquitous 380-500 nm band edge on Io may be unlikely to accumulate at Io's high resurfacing rates \citep{McEwenLunine1990comment}. Given the spectral similarity of all of the aforementioned species to Io's spectra (Figure \ref{fig:lab}) and the laboratory understanding of their formation and stability on the surface, it seems likely that all may be present in some amount, with an undetermined combination of S$_8$ and S$_\mu$ (perhaps akin to the quenched sulfur melts of \citet{MosesNash1991}) as the primary cause of the 380--500 nm absorption edge.

The identity of Io's 560 nm feature has also been a subject of debate. Suggestions have ranged from ferric iron \citep{NashFanale1977} to sodium sulfide \citep[Na$_2$S;][]{Nash1993, Gradie1984, NashNelson1979} to tetrasulfur \citep[S$_4$;][]{Nelson1978,Spencer1997, carlson2007io}. As mentioned by \citet{Spencer1997}, species containing ferric iron usually have prominent bands at longer wavelengths, which are not seen on Io. Between Na$_2$S and S$_4$, the former provides a better spectral match to Io's 560 nm absorption (Figure \ref{fig:lab}), as it is broader and at slightly longer wavelengths than the proposed band of S$_4$, which is nominally closer to 520 or 530 nm \citep[e.g.][]{Meyer1972}. However, though initially an attractive candidate as a possible source for the Na in Io's neutral clouds, Na$_2$S is now expected to be only a minor constituent of Io's volcanic emissions \citep{Fegley2000, Moses2002,SchaeferFegley2005}, which have since been shown to contain sufficient NaCl to explain the Na cloud \citep{Lellouch2003}. Even the more abundant NaCl likely only exists on the surface at an abundance of 1\% or less relative to SO$_2$ \citep{carlson2007io}, which likely explains why our HST spectra do not show the irradiation-induced NaCl color centers seen in equivalent observations of Europa \citep{TrumboEtAl2019, Trumbo2022}. Thus, S$_4$ seems the most likely of the discussed candidates for the 560 nm feature after consideration of our current understanding of Io's plume and atmospheric chemistry.  

In addition, it seems possible from a range of sulfur experiments that S$_4$ may variably affect the spectra of sulfur mixtures, potentially providing a better spectral fit to Io under different conditions, despite the apparently poor match to the 530 nm absorption of laboratory-prepared S$_4$ (Figure \ref{fig:lab}). Indeed, the spectrum of quenched red sulfur glass \citep[Figure \ref{fig:lab};][]{Meyer1972} and various spectra of ``purple", ``red", or ``brown" sulfur \citep[e.g.][]{Nelson1978, Gradie1980} show broader and shifted absorption that has also been attributed to the influence of S$_4$ in the samples. Molten sulfur \citep[e.g][]{Nelson1983} and some sulfur melt freezes \citep{MosesNash1991} also exhibit indistinct absorption at nearby wavelengths. In addition, S$_4$ is thought to be the likely source of the red color and strong absorption beyond 500 nm in laboratory samples of disulfur monoxide (S$_2$O; Figure \ref{fig:lab}), which has also been suggested as a possible Io surface species \citep{HapkeGraham1989, Hapke1989}. Therefore, while we have not seen a laboratory spectrum of S$_4$ or of an S$_4$-containing substance that, on its own, presents a satisfactory spectral match to Io's 560 nm band, we still find it a spectrally plausible species. 

Importantly, S$_4$ fits a conceptual model for the spatial distribution of the 560 nm absorption, which corresponds to the high latitudes and Pele's plume deposits \citep[Figure \ref{fig:560};][]{Spencer1997} and must therefore be tied to corresponding processes. Pele's plume contains gaseous S$_2$, which may result in the production of S$_4$ via photolysis and polymerization or annealing after deposition onto the surface \citep{Spencer2000io,carlson2007io}. S$_2$O is also an expected minor plume species, which may thermally decompose to produce S$_4$, as discussed by \citet{Spencer1997}. However, its expected low abundance relative to S$_2$ suggests that this mechanism would likely be a secondary source \citep{ZolotovFegley2000, carlson2007io}. Regardless, such direct production from continuously erupted plume gases may explain the specific association of the 560 nm band with Pele's ring, as well as the smaller diffuse red deposits seen near other volcanic centers in \textit{Galileo} imagery \citep{Geissler1999}.

S$_4$ is thermodynamically unstable at average Io surface temperatures \citep[e.g.][]{Meyer1972}, and therefore must be continuously produced to explain Io's 560 nm feature. The plume mechanisms discussed above fulfill this requirement for Pele's ring, but do not as directly explain the feature's association with the high latitudes. However, x-ray irradiation of S$_8$ can also lead to the production of S$_4$ \citep{Nelson1990}, and it is thus plausible that particle radiolysis would do the same, as has been observed in experiments with H$_2$S-bearing ices \citep{Mahjoub2017}. Though particle precipitation patterns on Io are complex and not entirely understood \citep{Paranicas2003io}, the low atmospheric density over its poles \citep[e.g][]{Strobel2001,Feaga2009io,Lellouch2015io} may facilitate enhanced high-latitude particle bombardment relative to the rest of the surface \citep{Johnson1997ganyio,Bagenal2020}, thereby providing a radiolytic source of S$_4$. In principle, precursors to its formation via this mechanism could also include alteration products of SO$_2$ that has condensed onto the poles (perhaps S or S$_2$O), which \citet{Wong1996} and \citet{Johnson1997ganyio} suggested could account for the darkened high latitudes. 

In addition to the possibility of enhanced radiolytic production near the poles, it seems plausible that their lower surface temperatures may also act to increase the preservation of unstable S$_4$, as posited by \citet{Spencer1997}. Based on this combination of potential factors, we agree with the interpretations of \citet{Spencer1997} and \citet{carlson2007io} in considering S$_4$ the best current candidate for Io's 560 nm absorption. However, we find this identification  tentative, given that laboratory experiments involving S$_4$ have yet to produce a satisfactory spectral match to the observed band.

\section{Conclusions} \label{sec:conclusions}
We present global, spatially resolved UV-visible spectra of the surface of Io from 200 to 1000 nm, which we obtained with HST/STIS. Using these data, we extract and map endmember components, obtain the geographies of observed spectral features, and discuss potential compositional insights by considering spectral comparison to laboratory samples as well as the context of Io' plume chemistry and surface processes. In agreement with past observations, our results indicate low-latitude deposits of SO$_2$, which have been stable across the roughly 40 years since the \textit{Voyager} flybys of Io. We suggest that future modeling work simultaneously examining the UV, visible, and infrared wavelengths may help resolve outstanding questions regarding the grain sizes and physical structures of SO$_2$ deposits across the surface. While we do not definitively identify Io's most prominent visible-wavelength absorptions---a deep 380--500 nm absorption edge and a prominent 560 nm band---we confirm their correspondence to the respective yellow and red materials investigated by \textit{Galileo} imagery and posit potential compositions. We agree with past interpretations that the 380--500 nm absorption edge likely reflects elemental sulfur, and suggest a composition featuring both S$_8$ and S$_\mu$ as physically and spectrally plausible. We also find our results to be consistent with past suggestions of the potential enrichment of Pele's pyroclastic ring and the high latitudes in the metastable short-chain allotrope S$_4$, which may form continuously from plume gases at Pele and possibly be independently produced via radiolytic processes and/or preserved by low temperatures toward the poles.

\acknowledgments
Based on observations made with the NASA/ESA Hubble Space Telescope, obtained at the Space Telescope Science Institute, which is operated by the Association of Universities for Research in Astronomy, Inc., under NASA contract NAS5-26555. These observations are associated with program $\#$15925. Support for program $\#$15925 was provided by NASA through a grant from the Space Telescope Science Institute, which is operated by the Association of Universities for Research in Astronomy, Inc., under NASA contract NAS5-26555. S.K.T. is supported by the Heising-Simons Foundation through a \textit{51 Pegasi b} postdoctoral fellowship. All of the data presented in this paper were obtained from the Mikulski Archive for Space Telescopes (MAST) at the Space Telescope Science Institute. The specific observations analyzed can be accessed via \dataset[10.17909/t9-5tsd-s772]{https://doi.org/10.17909/xd3d-wp93}. The authors thank Katherine de Kleer and Jonathan Lunine for useful discussions.
\newline

\facilities{HST(STIS)}
\software{astropy \citep{2013A&A...558A..33A}}

\end{document}